\documentclass[twocolumn]{iopart}

\usepackage{graphicx}
\usepackage{color}
\usepackage{iopams}

\begin{document}

\title{From Josephson junction metamaterials to tunable pseudo-cavities}

\author{D Zueco$^{1,2}$,  C
  Fern\'andez-Juez$^1$, J Yago$^1$, U. Naether$^1$,  B. Peropadre$^3$,
J J Garc\'{\i}a-Ripoll$^3$ and J J Mazo$^1$} 
\address{$^1$ Instituto de Ciencia de Materiales de Arag\'on y
  Departamento de F\'{\i}sica de la Materia Condensada,
  CSIC-Universidad de Zaragoza, E-50009 Zaragoza, Spain.}
\address{$^2$ Fundaci\'on ARAID, Paseo Mar\'{\i}a Agust\'{\i}n 36,
  50004 Zaragoza, Spain}
\address{$^3$ Instituto de F\'{\i}sica Fundamental, IFF-CSIC, Serrano
  113-bis, 28006 Madrid, Spain}
\ead{juanjo@unizar.es}

\begin{abstract}
  The scattering through a Josephson junction interrupting a
  superconducting line is revisited including power leakage.  We
  discuss also how to make tunable and  broadband resonant mirrors by
  concatenating junctions. As an application, we show how to construct
  cavities using these mirrors, thus connecting two research fields:
  JJ quantum metamaterials and coupled cavity arrays. We finish by
  discussing the first non-linear corrections to the scattering and
  their measurable effects.
\end{abstract}
\pacs{ 42.50.Dv, 
  03.65.Yz, 
  03.67.Lx, 
}
\submitto{\SUST}
\noindent{\it Keywords\/}: Josephson junction, scattering, open
transmission lines, superconducting cavities, metamaterials.
\maketitle
                             


\section{Introduction}

Superconducting quantum circuits have demonstrated to be a flexible
and scalable platform for quantum information processing
~\cite{You2011, Buluta2011}. Experimental proofs of simple quantum
algorithms~\cite{Dicarlo2009, Mariantoni2011}, the first steps for
many-body quantum simulations~\cite{Houck2012} and rapid improvements
in the coherence times pave the way towards more scalable designs and
experimental developments that beat what is classically computable or
simulable. In addition to the quantum information route, quantum
circuits are particularly interesting from a fundamental point of
view, as a platform where both matter and light can be accurately
engineered, with an unbeatable tunability, interaction
strength~\cite{Niemczyk2010,Forn-Diaz2010,Bourassa2009} and
scalability of designs.

While the experimental and theoretical focus was so far centered on
the qubits, a new type of experiments and proposals are beginning to
explore the study of propagating microwave photons and their
interaction with artificial matter (qubits) or Josephson
junctions. The building block of these studies is the scattering of
photons through a qubit~\cite{Shen2005, Zhou2008, Zhou2008a} or a
Josephson junction (JJ)~\cite{Zueco2012}. The reflection and
transmission properties of these nonlinear scatterers have been probed
in experiments with qubits \cite{Astafiev2010a, hoi2011, Astafiev2010}
and SQUIDs \cite{Jung2013}. More recently, arrangements of qubits or
JJs have been suggested to tailor the propagation of
light~\cite{Rakhmanov2008, Zagoskin2009, Nation2009, Hutter2011,
  Zueco2012, Mukhin2013}, conforming what is now called
\textit{quantum metamaterials}, the topic of this special issue. These
are low loss devices, since the underlying medium for the photon is a
superconductor at a very low temperature, but they introduce new
physics: from engineering of bandgaps and dispersion relation as in
classical wave propagation, to purely quantum effects such as
electromagnetic induced transparency (EIT)~\cite{Abdumalikov2010,
  hoi2011} and other quantum phenomena.

There is an alternative paradigm for the study and control of light in
quantum circuits and quantum optics: cavity-QED. Confining the light
in optical cavities or resonators, the photon-qubit coupling can be
enhanced (the Purcell effect) even to a point that the interaction
energy gets close to the photon or qubit energy \cite{Niemczyk2010,
  Forn-Diaz2010}. Without resorting to this enhanced interaction
regime, already at what is called the strong coupling, we find the
possibility of engineering new quasiparticles, the polaritons, which
are entangled states of light and
matter~\cite{hartmann06,Greentree2006,angelakis07}. These polaritons
can be engineered in arrays of microwave cavities with embedded
superconducting qubits~\cite{Koch2009, Hummer2012, Houck2012}, where
the combined matter-light excitations move freely and implement
sophisticated many-body Hamiltonians.

In this paper we merge both paradigms: that of quantum metamaterials
and that of polaritonic arrays. The basic idea is that a Josephson
junction may act as a mirror that reflects photons in a broad range of
frequencies. We will show that embedding these junctions in a
transmission line we can engineer localized modes that act as cavities
with a moderate quality factor. This is similar to other designs where
the pseudo-cavity is implemented by qubits~\cite{Chang2011a}, but we
extend the idea from a single cavity, with two junctions, all the way
to periodic arrangements of junctions that implement multiple coupled
cavities. We discuss when this image is valid, what are the effective
couplings between cavities and how these setups are related to our
previous proposals on quantum metamaterials~\cite{Zueco2012}.

The paper is organized as follows.  In the next section, Sect. \ref{sec:scatterer}, we review the
single junction scattering.  We include dissipation and a discussion
on the power leakage.  We complete the
study with the reflection/transmission characteristics for concatenated
junctions.
In section \ref{sec:cav} we apply the scattering theory for building cavities 
and relate the reflection and transmission of junctions with the
coupling between different cavities.  
We continue by  commenting on the first non linear corrections
(Sect. \ref{sec:nonlinear}) 
and the 
paper is finished with our conclusions.


\section{Josephson junctions as  scatterer}
\label{sec:scatterer}

JJs are present in almost any superconducting quantum circuit, because
they provide the non-linearity and the tunability which is needed in
those circuits. In particular, they make it possible to build qubits
and few-level systems~\cite{Makhlin2001}, or tunable linear and
non-linear
resonators~\cite{Castellanos-Beltran2008,Wilson2011,Ong2010}, and they
are also used to shape and enhance the qubit-resonator
coupling~\cite{Niemczyk2010}. In addition, junctions are at the heart of recent
works which introduce \textit{quantum metamaterials} for shaping the
transport of microwave photons~\cite{Rakhmanov2008, Zagoskin2009,
  Nation2009, Hutter2011, Zueco2012, Jung2013}. Working in the linear
regime, the JJs act as local scatterers that can form band gaps and
tailor the photon group velocity. Such setups may be generalized to
two dimensions developing, {\it e.g.}, left-hand
metamaterials~\cite{Zueco2012}. The main advantages of JJ-based
quantum metamaterials are that they can be tuned (replacing the JJs by
dc SQUIDs) and that their inherent losses are very small.

\begin{figure}
\includegraphics[width=0.6\columnwidth]{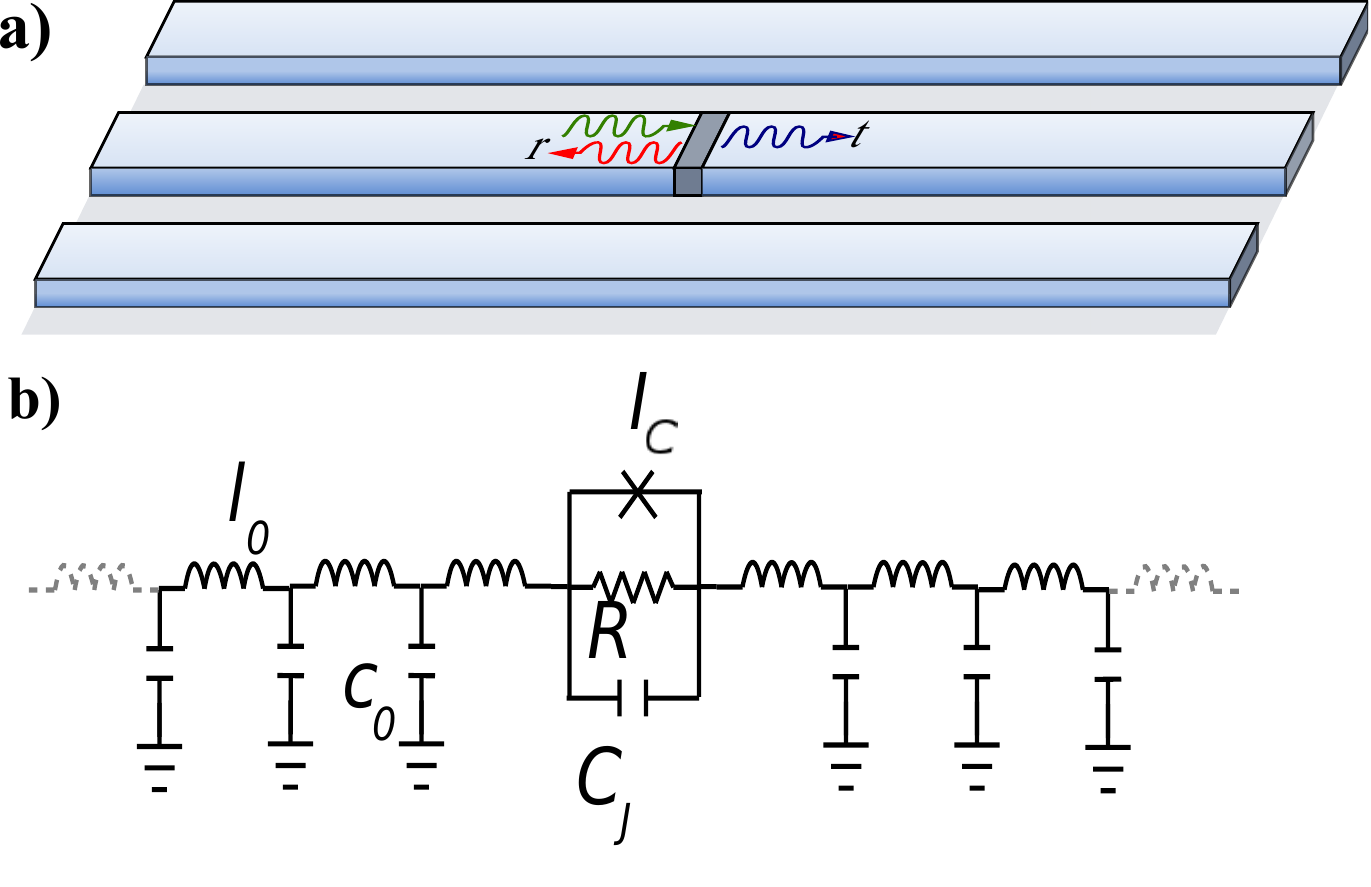}
  \caption{(color online) (a) An open transmission line interrupted by a Josephson junction (sketch). (b) Equivalent lumped element circuit, using the RCSJ model for the junction.}
\label{fig:jjtl}
\end{figure}

The minimal setup for observing the scattering of photons through a JJ
consists on transmission line interrupted by a single junction, as
sketched in Fig.~\ref{fig:jjtl}(a). Constructing the equivalent
lumped-element circuit from Fig.~\ref{fig:jjtl}(b) and taking the
continuum limit provides us with a Lagrangian that describes a
transmission line with the junction~\cite{Niemczyk2010, Ong2010,
  Bourassa2009, Zueco2012}
\begin{eqnarray}
\label{tl-jj}
\mathcal L &= \frac{1}{2} \int_{-\infty}^{0_-} {\rm d}x  \left[ c_0 (\partial_t \phi) ^2 - \frac{1}{l_0} (\partial_x \phi) ^2\right]
\\ \nonumber
&+ \frac{1}{2}  \left (\frac{\Phi_0}{2 \pi} \right )^2 C_J \left (\frac{d \varphi}{dt} \right) ^2
-\left (\frac{\Phi_0}{2 \pi} \right ) I_C \cos{\varphi}
\\ \nonumber
&+ \frac{1}{2} \int_{0_+}^{\infty} {\rm d}x \left[ c_0 (\partial_t \phi) ^2 - \frac{1}{l_0} (\partial_x \phi) ^2\right].
\end{eqnarray}
Through this work we assume that the capacitance and inductance per
unit of length, $c_0$ and $l_0$ respectively, are constant along the
line. The junction, placed at the origin, $x=0$, is characterized by a
capacitance $C_J$ and a critical current $I_C$. While the microwave
transmission line is described by a continuous field denoting the
flux, $\phi(x,t)$, the Josephson junction has a unique degree of
freedom associated to it, which is the gauge invariant phase,
$\varphi$. This is given by
\begin{equation}
\label{varphi}
\varphi = \Delta \theta - \frac{2 \pi}{\Phi_0} \int_{0_-}^{0_+} {\bf A} ({\bf r},t) \cdot {\rm d} {\bf l} \; ,
\end{equation}
where $\Delta \theta$ is the superconducting phase difference and ${\bf A}({\bf r},t)$ is the vector potential. Note that since the junction is connected to two semi-infinite transmission lines, $\varphi$ is not be an independent variable, but depends on the incoming and outgoing fluxes
\begin{equation}
\varphi (t) \propto (\phi(0^+,t) - \phi(0^-,t)).
\end{equation}

For studying the scattering of photons in the linear regime we must consider only perturbations of the equilibrium situation. We thus introduce the changes of the field, $\tilde \phi (x,t)$, with respect to the static background flux $\phi^{(0)} (x)$,
\begin{equation}
 \phi (x,t) = \phi^{(0)} (x) + \tilde \phi (x,t).
\end{equation}
We do the same for the junction, introducing a flux variable $\delta \phi (t)$ associated to the time fluctuations for the flux across the junction
\begin{equation}
\label{wtvp}
\varphi (t) = \varphi^{(0)}+\frac{2\pi}{\Phi_0} \delta \phi (t)
\end{equation}
Here $\varphi^{(0)}$ stands for the equilibrium solution for the phase and $V = (\Phi_0/2\pi) \dot \varphi = \dot{\delta \phi}$  is the expected voltage-flux relation.

Using the previous variables and the Lagrangian formalism we can
construct equations for the propagation of photons through the
junction~\cite{Zueco2012}. We will complement those conservative
differential equations with a model for the (possibly weak) losses on
the junction itself. This is done in the lumped-element circuit model
introducing the RCSJ model for the junction~\cite{Orlando}: a
resistance in parallel with the junction capacitance and the nonlinear
inductance.

For the study of few-photon scattering, out of the resulting
equations, we only need to consider the ones that match the fields to
the right and to the left of the junction, through the Josephson
relation. This is the equation of current conservation, which reads
\begin{equation}
\frac{1}{l_0} \partial_x \phi(0_-, t) =
\frac{\Phi_0}{2 \pi} C_J \ddot {\varphi} + I_C \sin{\varphi}
+\frac{1}{R} \frac{\Phi_0}{2 \pi} \frac{d \varphi}{dt} = \frac{1}{l_0} \partial_x \phi(0_+, t) 
\label{wtcc}
\end{equation}
Since our studies focus on the low-power (few-photon) regime, we can
linearize the equations, assuming small fluctuations in the junction
phase, to obtain
\begin{equation}
\label{scjjlf}
\frac{1}{l_0} \partial_x \tilde \phi(0_\pm, t) = 
C_J  \, \ddot {\delta  \phi} + \frac{1}{L_J} \, \delta \phi 
+\frac{1}{R} \dot {\delta  \phi} 
\end{equation}
with $L_J=\Phi_0/(2\pi I_C \cos{(\varphi^{(0)})})$. Note that the
linearization also provides us with the static configuration for the
flux: $1/l_0 \partial_x \phi^{(0)} (x) = I_c \sin(\varphi^{(0)})$.

In the linearized theory, the stationary scattering solutions can be
written as a combination of incident, reflected and transmitted plane
waves:
\begin{equation}
\label{ansatz_jj}
\tilde \phi(x,t) =
A_\phi
\left \{
\begin{array} {cc}
{\rm e}^{i ( k x - \omega t)} + r {\rm e}^{-i ( k x + \omega t)} & (x
< 0),
\\
t {\rm e}^{i ( k x - \omega t)} & (x > 0),
\end{array}
\right .
\end{equation}
where $A_\phi$ is the field amplitude, and $r$ and $t$ are the
reflection and transmission coefficients, respectively. Outside the
junctions the incoming and outgoing photons follow a linear dispersion
relation, $\omega=v k$ with $v=1/\sqrt{l_0c_0}$, which we can
substitute in the previous ansatz, to compute the transmission and
reflection
\begin{equation}
\label{r-jj}
r
=
\frac{1}{1 - i 2 
z \frac{1}{\bar \omega}
\left ( 
\bar \omega ^2
+ i \gamma \bar \omega
-1
\right )}
,\quad
t = 1 -r.
\end{equation}
The scattering properties are a function of the photon frequency,
$\bar \omega = \omega / \omega_p$, rescaled with the Josephson plasma
frequency $\omega_p = 1/\sqrt{L_J C_J}$. They also depend on the
impedance of the line, $Z_0=\sqrt{l_0/c_0}$, and the junction,
$Z_J=\sqrt{L_J/C_J}$, through their ratio $z= Z_0/Z_J$. Finally,
dissipation enters via the dimensionless parameter $\gamma = Z_J / R$.

\begin{figure}
\includegraphics[width=1.\columnwidth]{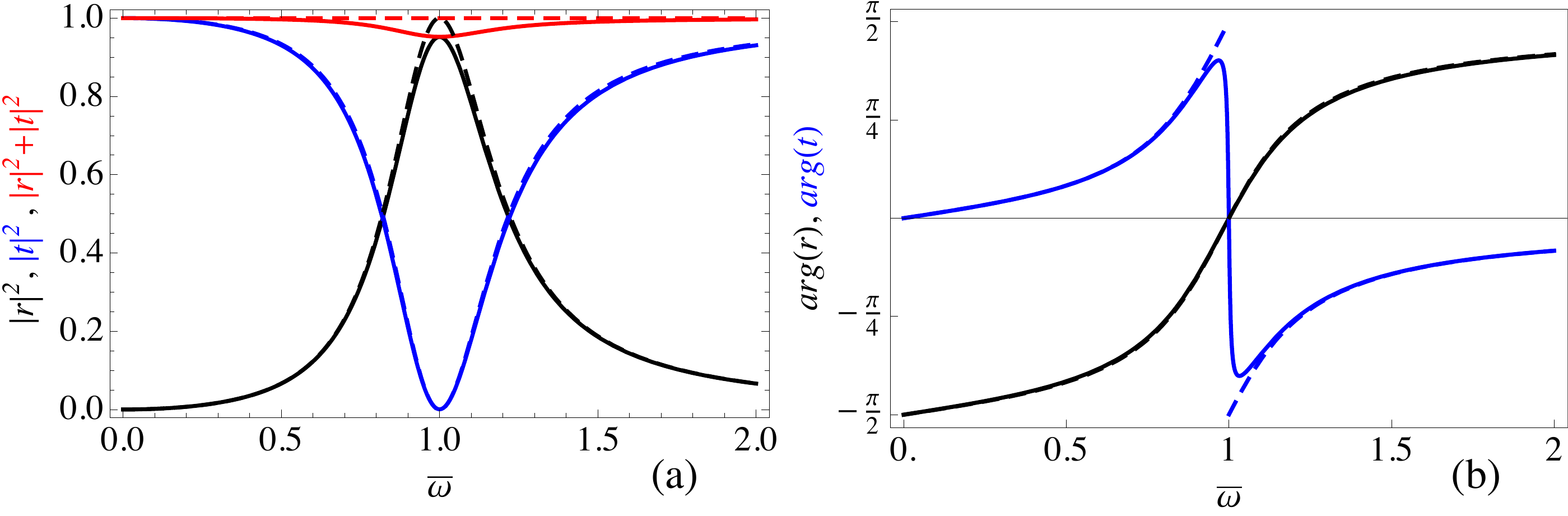}
  \caption{(color online) Transmission and reflection properties of a
    single junction acting as a scattering element. (a) Transmission,
    $t$, and reflection, $r$, for a junction without losses (dashed)
    and with losses ($\gamma=0.01$ or $R\sim 5 \times 10^3\,\Omega$,
    solid), using $z=1.25$. We also plot the total outgoing photons,
    $|r|^2+|t|^2$, which close to resonance is less than 1 when in
    the presence of losses. (b) Phase of the transmitted and
    reflected photons, $arg(r)$ and $arg(t)$, respectively. Note the
    phase jump close to reflection, which is damped by the losses.}
\label{fig:jj-sc}
\end{figure}

In the previous formulas we find a resonance at the plasma frequency
of the junction, $\bar\omega=1$ or $\omega=\omega_p$. At this point
the reflection becomes maximum and, in absence of dissipation, the
junction behaves as a perfect mirror. A similar resonance mechanism is
found when studying the scattering of photons through qubits
~\cite{Shen2005,Astafiev2010a}. This is illustrated in
figure~\ref{fig:jj-sc}, where we plot the reflection-transmission
characteristics, in the ideal (dashed) and dissipative (solid) cases.
Note also that, as in classical wave propagation, the maximum of the
reflection is accompanied by a phase jump of the scattered photon
across the junction, as shown in figure~\ref{fig:jj-sc}b.

\subsection{Power leakage}

We already saw in Fig.~\ref{fig:jj-sc}(a) that for a dissipative
junction the transmitted and reflected powers do not add up to
one. Instead, some photons are absorbed by the junction and get lost,
distorting slightly the resonance. It is instructive to have a closer
look at the dissipation. For that we study the energy function in an
interval of the line containing the junction:
\begin{eqnarray}
E(x_0)
=&
\frac{1}{2}
\int_{-x_0}^{0_-} {\rm d}x 
\left[
c_0 (\partial_t \phi) ^2 +
\frac{1}{l_0} (\partial_x \phi) ^2\right]
\\ \nonumber
&+
\frac{1}{2}  \left (\frac{\Phi_0}{2 \pi} \right )^2 C_J \left (\frac{d \varphi}{dt} \right) ^2
+\left (\frac{\Phi_0}{2 \pi} \right )
I_C \cos{\varphi}
\\ \nonumber
&+
\frac{1}{2}
\int_{0_+}^{x_0} {\rm d}x 
\left[
c_0 (\partial_t \phi) ^2 +
\frac{1}{l_0} (\partial_x \phi) ^2\right]
\,.
\end{eqnarray}
Differentiating this functional with respect to time one obtains an
expression that relates the dissipated power (loss of energy) to the
flux drop around the junction
\begin{equation}
\label{power}
P = \frac{\partial E (x_0)} {\partial t} = \frac{1}{l(x)} (\partial_x \phi)  \dot \phi (x) |_{-x_0}^{x_0}.
\end{equation}
Rather than the instantaneous power it is more illustrative to
compute the average power in a cycle of the photon oscillations,
\begin{equation}
\label{barP}
  \bar P = \frac{\omega}{2 \pi} \int _0 ^{2 \pi /\omega}  {\rm d} \tau P (\tau).
\end{equation}
Combining the {\it ansatz} (\ref{ansatz_jj}) together with the
formulas for the transmission and reflection, $t$ and $r$ in
Eq.~(\ref{r-jj}), we get an expression that it is intuitively
appealing
\begin{equation}
\label{power1}
\bar P
=
A_\phi^2
\frac{\omega^2}{2 Z_0}
\Big (
1
-
|r|^2
-
|t|^2
\Big )
=
2 A_\phi^2  \, Z_0 \gamma \, \omega^2 |r|^2
\end{equation}
Roughly, the dissipated power is proportional to the incoming
intensity and the reflected power, but also to the effective
dissipation rate $\gamma$ of the junction\footnote{It is worth
  mentioning that the same result is obtained by simply using the
  simple formula for the power dissipation in a junction $P = V^2/R =
  \dot{\delta \phi}^2/R$~\cite{Orlando}.}. Dissipation is thus maximum
only close to resonance, $|r| \sim 1$, where it degrades the
reflection due to photon loss. Since the losses do not shift the
resonance, which remains fixed at the condition $ \bar \omega = 1$, we
obtain that the maximum dissipated power is
\begin{equation}
|r|^2_{(\bar \omega = 1)}=
\frac{1}{(1 + 2 z \gamma)^2}
\;
\longrightarrow
\bar P _ {(\bar \omega = 1)}
=
 \frac{A_\phi^2}{Z_0}   \frac{2 z \gamma}{ (1 + 2 z\gamma)^2}.
\end{equation}
Note that these results also apply to the context of microwave
photodetection, where the loss mechanism is not the junction but an
imperfect qubit. In this situation the maximum dissipated power can be
directly related to the maximum detection
efficiency~\cite{Romero2009}.

\subsection{JJs as tunable mirrors}
\label{sec:tunable}

While JJs may act as mirrors, it is well known that capacitors or cuts
in the transmission line are higher quality mirrors. Why then study
metamaterials constructed of JJs? There are several answers to this
question. For starters, JJs offer the potential for nonlinear
scattering of photons, discussed in Sect.~\ref{sec:nonlinear}. Most
important, the reflectivity and transmission of a junction are
frequency-dependent, acting as filters which suddenly become tunable
when, instead of a junction, we use a dc SQUID. A dc SQUID is a circuit
consisting on two junctions connected in parallel and threaded by some
magnetic flux. The scattering properties of the dc SQUID are the same as
those of a junction, but with the advantage that the plasma frequency,
$\omega_p = 1/\sqrt{L_J C_J}$, can now be tailored {\it in situ} via
that external magnetic flux, $L_J=\Phi_0/[4\pi I_C \cos(\pi \Phi_{\rm
    ext}/\Phi_0)]$, with $\Phi_{\rm ext}$ the external flux through
the SQUID [Cf. Eq. (\ref{scjjlf})].

Another way in which the reflectivity of the junctions can be tuned is
by combining several of them. As we will see below, this changes the
bandwidth of the JJ filter, modifying the frequency-dependent
transmission and reflection, and also the effect of losses ---the
mirror becomes more perfect. The idea of combining multiple scatterers
to study their collective properties has been explored in the context
of qubits, where two-level systems act as frequency-dependent
mirrors~\cite{Chang2011a}. Inspired by this work, we follow a similar
route.

\begin{figure}
  \includegraphics[width=0.5\linewidth]{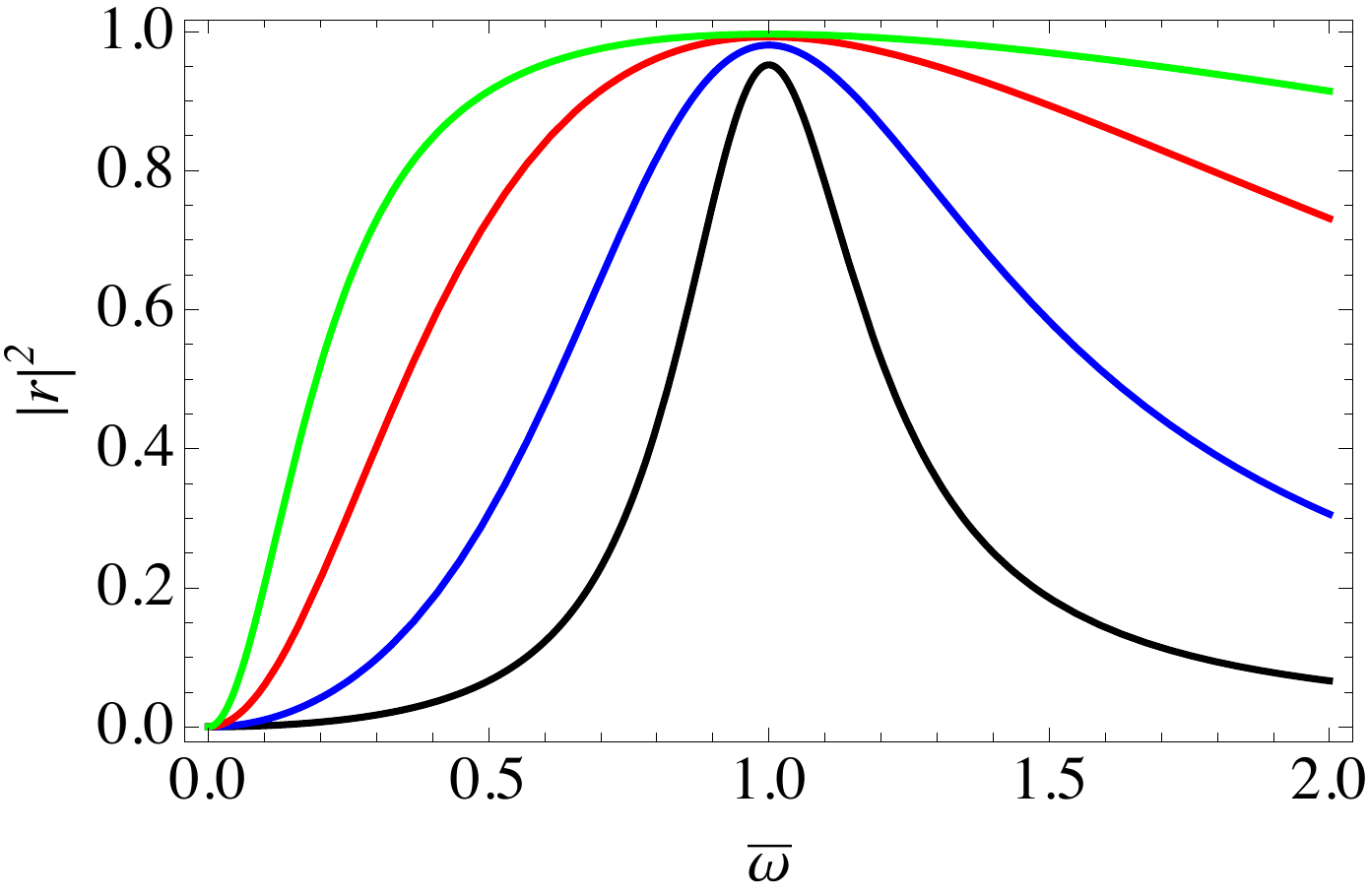}
  \caption{ $|r|^2$ as a function of the number of junctions $N$. From
    bottom to top $N=1, 2, 4$ and 8 (colored as black, blue, red and
    green respectively). The rest of the parameters like in figure
    \ref{fig:jj-sc}.  }
\label{fig:njj}
\end{figure}

We will consider an arrangement of $N$ junctions disposed one after
another separated by a distance $d$ in the same transmission line. If
$d$ is large, we expect large oscillations of the scattering
properties due to the constructive and destructive interference of the
photons that are multiply reflected by the different
junctions~\cite{Romero2009,Chang2011a}. Instead, we will focus on the
limit in which the separation among junctions is so small, $d \ll
\lambda$, that we can take the limit $d=0$ in the formulas. This is a
very realistic assumption due to the difference in size ---$d$ lays
around the nanometer, while $\lambda$ is around the millimeter to
centimeter. In this limit, we can neglect the phase acquired by the
photons due to their propagation and regard the $N$ junctions in
series through a generalization of Eq.~(\ref{scjjlf}) to $N+1$
equations of the form
\begin{equation}
\label{cNjj}
\frac{1}{l_0} \partial_x \tilde \phi(0_\pm, t) = 
C_J  \, \ddot {\delta  \phi}_n + \frac{1}{L_J} \, \delta \phi_n 
+\frac{1}{R} \dot {\delta  \phi}_n 
\; \quad n = 1, ..., N
\end{equation}
where $\delta \phi_n$ are the flux fluctuation across each
junction. Assuming for simplicity that all junctions are identical, we
obtain that the total flux fluctuation $\delta\phi = \tilde \phi (x_+)
- \tilde \phi (x_-)$ is equally spread among junctions $\delta \phi_n
= \delta \phi_m= \delta \phi / N$. In other words, the voltage drop
along the circuit is equally divided among the junctions and the
formula for the reflection gets then modified accordingly:
\begin{equation}
\label{r-njj}
r = \frac{1}{1 - i \frac{2 z}{N}  \frac{1}{\bar \omega} \left ( \bar \omega ^2 + i \gamma \bar \omega -1 \right )},\quad t = 1 -r.
\end{equation}
The $N$ junctions behave as a single circuit, where transmission and
reflection coefficients are modified by a factor $1/N$ in the
denominator. The result is that, for a fixed ratio $Z_J/Z_0$ the
number of concatenated junctions enlarge the resonance, as shown in
Fig.~\ref{fig:njj}.

The concatenation of junctions also has a remarkable effect on the
dissipation, increasing the quality factor of the
mirror. Qualitatively, we expect that since the voltage drop is
equally divided among the $N$, junctions, $\dot {\delta \phi}_n = \dot
{\delta \phi} \sim r$, the losses at each junction should be decreased
by the corresponding factor $1/N^2$.  Summing over all $N$ junctions,
the result should be that the dissipation scales as $1/N$, being
reduced and increasing the quality of the overall circuit. Moreover,
from Eq.~(\ref{r-njj}) a reduction in the losses should be accompanied
by an increase of the reflectivity.  We can confirm this line of
thought, combining all previous formulas for the reflection and the
dissipated power
\begin{equation}
|r|^2_{\bar \omega = 1}=
\frac{1}{\left (1 +\frac{2z}{N} \gamma \right)^2}
\;
\longrightarrow
\bar P _{\bar \omega = 1}
=
\frac{A_\phi^2}{Z_0} \frac{ \frac{2z}{N} \gamma}{(1+ \frac{2z}{N} \gamma)^2}
\end{equation}
It would thus seem advantageous to use arrangements of multiple
junctions or SQUIDs to build cavities, not only because the quality
factor increases, but because the mirrors become broadband, allowing
for a better definition of the localized mode. This is indeed the
topic of the following section.

\section{Pseudo-cavities}
\label{sec:cav}

We have seen that Josephson junctions act as perfect mirrors for broad
ranges of frequencies. It is natural then to wonder whether such
mirrors can be used for building microwave cavities and what are their
properties: quality factor, wavelengths, etc. We will show how these
questions can be addressed using the scattering formalism developed
above, studying the formation of one cavity, the coupling of two
cavities and how these setups can be optimized and scaled up.

\subsection{Localized mode for two junctions}

\begin{figure}[t]
  \centering
  \includegraphics[width=1.\linewidth]{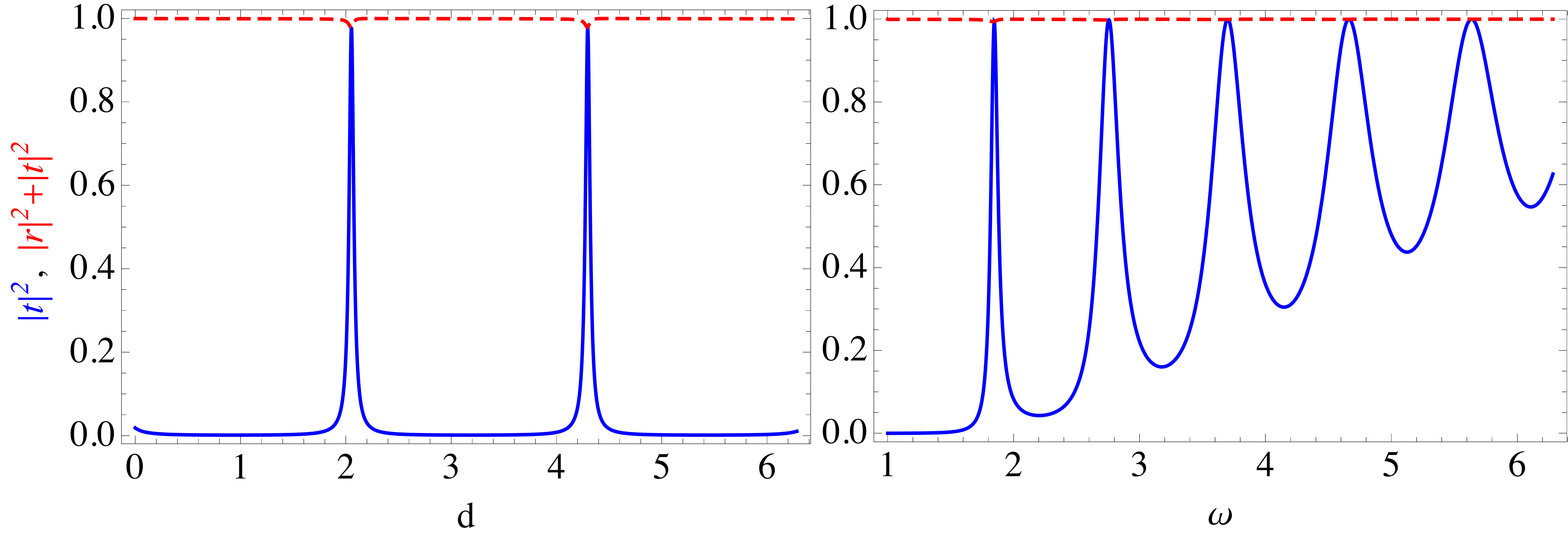}%
  \caption{Transmission of a setup with two junctions separated a
    variable distance $d$ (blue line), and the sum of the transmission and
    reflection (dashed red line). Left: Resonances as a function of the
    separation between junctions, for $\omega=1.4$ and $z=0.2,
    \gamma=0.01$. Right: For a fixed cavity separation, $d=\pi$,
    different resonances associated to the fundamental and first
    harmonics of the pseudo-cavity.}
  \label{fig:one-cavity}
\end{figure}

Let us consider a photon propagating through a setup consisting on two
junctions separated by a distance $d$. The transfer matrix of a single
junction has a simple form in terms of the transmission and reflection
coefficients
\begin{equation}
  T = \left(\begin{array}{cc}
      1-\frac{r}{t} & \frac{r}{t} \\ -\frac{r}{t} & \frac{1}{t}
      \end{array}\right).
\end{equation}
This matrix connects the left- and right-propagating components ($a$
and $b$ below) of a wave to one side and another side of the junction:
\begin{equation}
  \left(\begin{array}{c} a_L \\ b_L\end{array}\right)
  = T
  \left(\begin{array}{c} a_R \\ b_R\end{array}\right).
\end{equation}
For the ordinary propagation of a photon we have a similar transfer
matrix, which only adds a phase to the fields
\begin{equation}
  T_{prop}(d) = \left(\begin{array}{cc}
      e^{i \omega d /c} & 0 \\ 0 & e^{-i \omega d/c}
      \end{array}\right),
\end{equation}
where $c$ is the speed of the photons and $\omega$ the frequency. The
whole transfer matrix of this cavity-like setup is
\begin{equation}
  T_2 = T T_{prop}(d) T,
\end{equation}
which has an associated reflection coefficient
\begin{equation}
  r_2 \propto 2 z (\omega^2 + i \gamma \omega - 1) \cos(d\omega/c) +
  \omega \sin(d\omega/c).
\end{equation}
If we neglect the dissipation, the reflection has minima on a regular
set of points, given by the equation
\begin{equation}
  \tan(d\omega/c) = \frac{2z}{\omega}(1-\omega^2). 
\end{equation}
All the minima are spaced a distance $\pi c / \omega$, but their basic
frequency is not exactly the wavelength of the photon, as shown in
Fig.~\ref{fig:one-cavity}. Note also that for each separation between
junctions, $d$, we also have a large number of resonances,
corresponding to the bare resonance and the first harmonics. These
additional modes are broader and have stronger decay rates, as it
happens with similar setups where the mirrors are implemented using
qubits instead of junctions~\cite{Dong2009}.  In fact, for a
sufficiently broadband mirror (assume $r={\rm const.}$ in the
frequencies of interest) and high reflectivity, the quality factor is
given by \cite{Hummer2013}
\begin{equation}
Q=\frac{\omega d} {2 v_g ( 1 - |r|^2)}
\end{equation}
with $v_g$ the group velocity in the line.  Therefore and as expected,
$Q$ increases with $N$.  Note that the leakage in the cavity can be minimized by
putting more junctions and it can be adjusted by detuning some of the
junctions that act as a mirror.

\subsection{Coupled cavities}

\begin{figure}[t]
  \includegraphics[width=1.\linewidth]{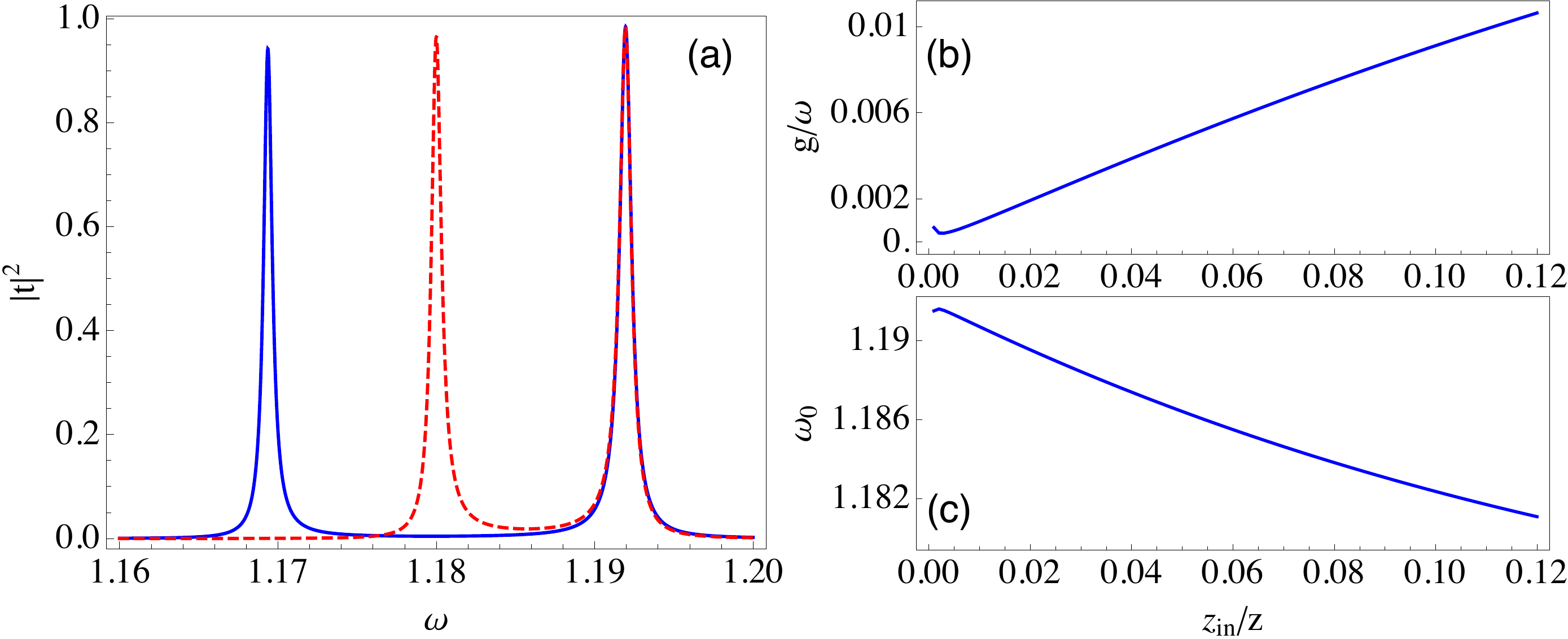}%
  \caption{Transmission of a setup with two pseudo-cavities separated
    by an intermediate junction with parameter $z_{in}$. The
    cross-talk between the localized modes of both cavities produces
    two separated peaks in the transmission, whose separation is
    proportional to the coupling strength. (a) Resonances for
    $z_{in}=0.1$ (solid blue) and $z_{in}=0.05$ (dashed red). (b)
    Effective coupling strength as a function of $z_{in}/z$.  (c) The
    central frequency between the peaks, $\omega_0$, shifts because of
    an interaction-induced renormalization. All simulations are
    computed assuming $\gamma=0.0001$, $d=2.6$ for the interjunction
    distance and using $z=0.1$ for the outer junctions.}
  \label{fig:coupled}
\end{figure}

To scale up the idea of implementing localized photons using
junctions, we need to study the effective coupling between two such
localized modes. For that we assume now a setup that consists on three
junctions. The middle junction, characterized by $z_{in}$
[c.f. Eq.~(9)], acts as a coupling element, while the outer junctions,
characterized by $z$, separate the two cavities from the outer
world. As before, we expect to have resonances at frequencies matching
those of a system of coupled cavities, which we find studying the
transmission and reflection properties of the combined setup.

In Fig. 5(a) we plot those properties for a setup with three
junctions.  We observe the appearance of two peaks around the central
frequency $\omega \cong 1.18$.  These two peaks account for the
cavity-cavity coupling. As usual we estimate this effective coupling
by using
\begin{equation}
  \omega_{\pm} = \omega_0(g) \pm \frac{g}{2},
\end{equation}
where the $g$ is the coupling strength and the middle frequency,
$\omega_0$, may be slightly renormalized due also to the
interactions. In Fig.~\ref{fig:coupled}(a) we plot the transmission
for two different $z_{in}$, given different couplings and central
frequencies.  A full dependence of the latter are shown in
Figs.~\ref{fig:coupled}(b) and (c). In particular, in
Fig.~\ref{fig:coupled}(b), we see that the interaction grows linearly
with the ratio $z_{in}/z$. This was expected: decreasing
$z_{in}$ implies increasing the reflectivity of the middle junction for
this particular frequency, thus decoupling nearby photons. Finally, as
the peak at the highest frequency remains fixed, the central frequency
also moves to higher frequencies when the cavity-cavity coupling is
enhanced as plotted in Fig.~\ref{fig:coupled}(c).

\subsection{Arrays of cavities}

The previous idea can be scaled up to construct arrays of coupled
cavities. We simply have to use more junctions spaced regularly. The
study of such systems becomes actually much simpler in the limit of
infinite many junctions, when we focus on the solutions that preserve
the number of photons, $\gamma=0$. In that case we can study the
eigenstates of the problem assuming translational invariance, that is
$\psi(x+d) = \psi(x) \exp(\pm ik)$, with $k$ the quasimomentum. As
explained in Ref.~\cite{Zueco2012}, we have to solve the problem
\begin{equation}
  2\cos(k) = \mathrm{tr}\left[ T\,T_{prop}(d)\right],
\end{equation}
which in our case reads
\begin{equation}
  \cos(k) = \cos(d\omega) + \frac{\omega \sin(d\omega)}{2z(\omega^2 + i\gamma\omega  - 1)},
\end{equation}
where we will assume $\gamma=0$ for simplicity.

\begin{figure}[t]
  \includegraphics[width=0.55\linewidth]{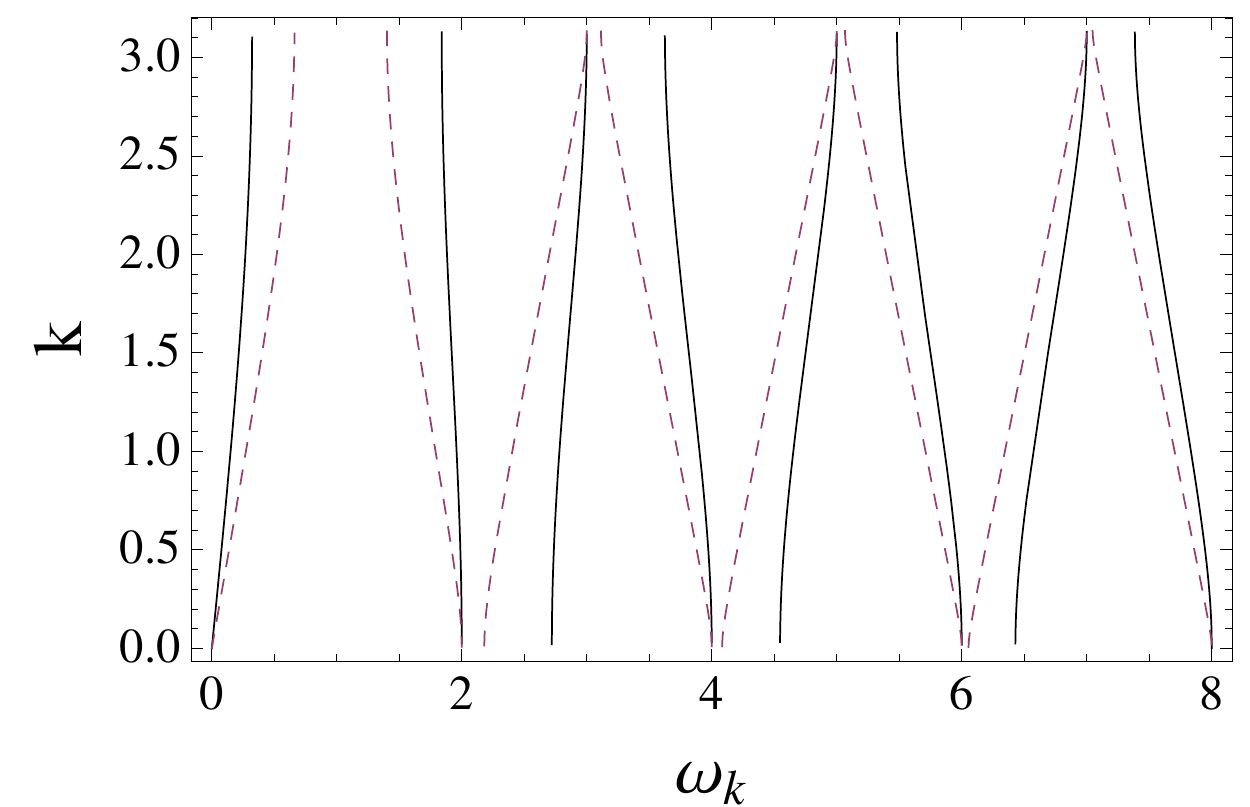}%
  \caption{Eigenenergies $\omega_k$ vs. lattice quasimomentum $k$,  for an array of junctions separated a distance $d=\pi$, with $z=0.1$ (solid) and $z=1$ (dashed) .}
  \label{fig:cavity-array}
\end{figure}

In our previous work~\cite{Zueco2012} we focused on the creation of
bandgaps and chose a junction separation, $d=0.3$, such that the
photons were in the region with transmission close to one. In this
work we are interested in recreating localized modes and thus $d$ must
be a distance that resonates with the junction. Thus, the value of $d$
should be close to an integer multiple of $\pi$, producing plots such
as the one in Fig.~\ref{fig:cavity-array}. In those plots we
appreciate the existence of an infinite series of bands with a width,
$\Delta\omega_n$, that grows with the frequency. We can regard these
bands as the results of photons hopping between the
pseudo-cavities. The central frequency of the $n$-th band would then
correspond to the $n$-th harmonic of the pseudo-cavity
[cf.~Fig.~\ref{fig:coupled}], while the width of the cavity will be
related to the cavity-cavity coupling, $g$, through the usual relation
in a tight-binding model, $g=\frac{1}{2}\Delta\omega_n$.

Naturally, the tight-binding approximation will work better when (i)
the bands are mostly flat and (ii) the gap between consecutive bands
remains large. This is, in turn, related to the properties of the
junction that we use to build the cavities: the bandwidth decreases
and the gap increases as we make $z$ smaller
[cf.~Fig.~\ref{fig:cavity-array}]. If we ensure the limit in which the
localized-mode approximation remains valid, we can use this
coupled-cavity array as a basis for the study of polariton
physics~\cite{hartmann06,Greentree2006,angelakis07}, for instance,
bridging the gap between the study of JJ quantum metamaterials and the
quantum simulation of many-body physics.

\section{Non-Linear corrections for the scattering}
\label{sec:nonlinear}

So far, we have assumed only linear dynamics for the junction. It
behaved as a local impurity in the line with a resonance frequency
$\omega_p$ and dissipation $\gamma$.
In fact all the applications discussed were rooted in the linear
approximation.
On the other hand, the JJ is the paradigm of a nonlinear system,
ranging from applications in classical dynamics: the circuit analogue
for the pendulum, and in quantum physics: the artificial atom or
qubit.
Therefore it seems reasonable to check up to what extent the linear
theory is accurate.
%
%
%
Since we are interested in an estimation of the appearance of
nonlinear corrections we  fix our attention to the simplest situation,
the scattering through a single junction neglecting dissipation.

%
%

For accounting to the nonlinear scattering we proceed as usual in
nonlinear optics and perform an  harmonic expansion for the left and
right fields \cite{Boyd2003}: 
\begin{equation}
\phi (x,t)
=
A_\phi
\left \{
\begin{array} {cc}
{\rm  Re} \left [ {\rm e}^{-i (kx -\omega t)}
+
\sum_n r_n  {\rm e}^{i n (kx +\omega t)}
\right ]
 & (x
< 0)
\\
{\rm Re} \left [
\sum_n
t_n
 {\rm e}^{-i n (kx -\omega t)}
\right ]& (x > 0)
\end{array}
\right .
\label{annl}
\end{equation} 
We emphasize here that we must work with the real part from the
beginning, since in the nonlinear case we will face with product
between different harmonics.  For the following it is convenient to
split the reflection and transmission coefficients in real and
imaginary parts $r_n = r_n^\prime + i r_n^{\prime \prime}$ and $t_n =
t_n^\prime + i t_n^{\prime \prime}$.  We replace this {\it ansatz} in
the equations for the current conservation (\ref{wtcc}). The equality
of the current at both sides of the junction, $1/l_0
\partial_x \tilde \phi (0_-, t) = 1/l_0\partial_x \tilde \phi (0_+,
t)$, implies
\begin{equation}
r_n^\prime + t_n^\prime
=
\delta_{n,1}
\;
\quad
r_n^{\prime \prime}=
t_n^{\prime \prime}
\, .
\end{equation}
Next, we match this current with current through the junction computed
expanding the sine function up to the fifth order:
\begin{eqnarray}
1/l_0 \partial_x \tilde \phi (0_-,
t) &=& 
C_J  \, \ddot {\delta  \phi} + 1/L_j\delta \phi  \\ \nonumber
&-&1/3! (2 \pi)^2/ \Phi_0^2L_J  \delta
\phi ^3 + 1/5! (2 \pi)^4/ \Phi_0^4L_J
\phi ^5.
\end{eqnarray}  Then we merge this expansion with the one for the fields
in Eq. (\ref{annl}) and match the different harmonics (fifth order).

\begin{figure}
\includegraphics[width=1.2\columnwidth]{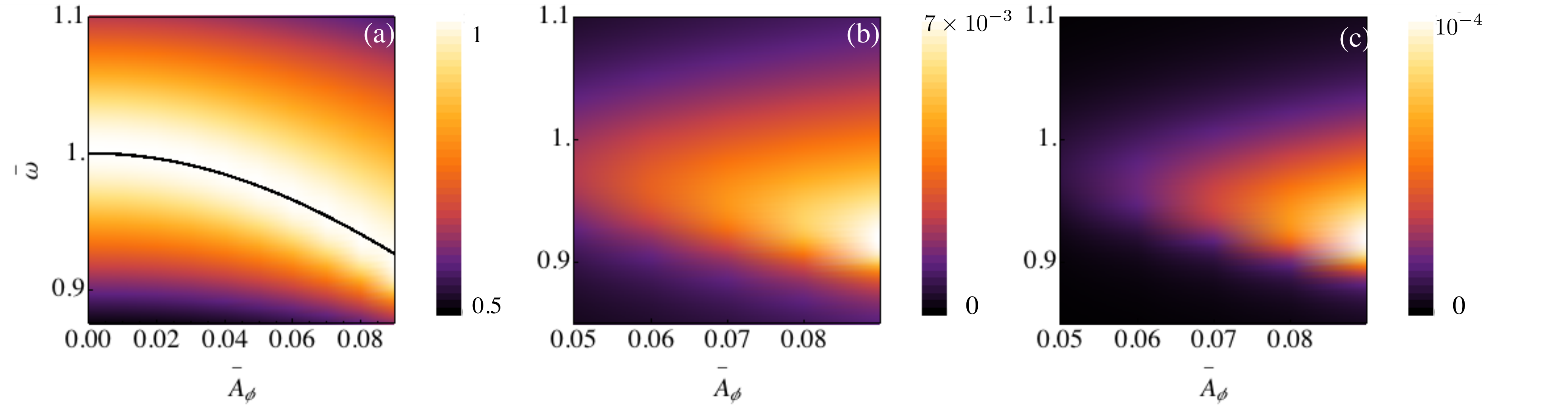}
  \caption{ $|r_1|^2$, $|r_3|^2$ and $|r_5|^2$ as a function of the
    incoming amplitude $\bar A_\phi = A_\phi / \Phi_0$ and the
    normalized frequency $\bar \omega = \omega / \omega_p$.
    Black line is the theoretical prediction (\ref{Tpend}) }
\label{fig:r-nl}
\end{figure}

The numerical results are summarized in the figure \ref{fig:r-nl}. We
notice that the nonlinear effects increase with the amplitude field
$A_\phi$ and that they are present around $\omega_p$.  We first plot
$|r_1|$ (the reflected photons with the same frequency). The graphics
show the expected shift in the resonance frequency $\omega_p$. The
resonance frequency moves to lower frequencies. This is a consequence
of the increase (decrease) of the pendulum period (frequency) as the
amplitude increases. An analytical formula is available in textbooks:
\begin{eqnarray}
\label{Tpend}
T &=& \frac{4}{\omega_J}
\int^{\pi/2}_0
\frac {{\rm d} \theta}
{\sqrt{1 - 
(2 \pi \bar A_\phi)^2 \sin (x)^2}}
\\ \nonumber
&\cong&
2 \pi \Big (
1+
\frac{1}{4}
( 2 \pi \bar A_\phi)^2
+
\frac{3}{4}
( 2 \pi \bar A_\phi)^4
\Big )
\,
\end{eqnarray}
where we have introduced $\bar A_\phi = A_\phi /\Phi_0$.
This formula is plotted together with the numerical results, showing
the agreement.

The other two plots stand for the $|r_3|$ and $|r_5|$, providing the 
third and fifth harmonic generation. 
These parameters, albeit small, are maximal whenever $|r| \cong 1$.
To understand the latter, we remind that the current through the JJ
depends on the flux difference $\delta \tilde \phi \sim t -r -1 =-2r$
entering in the sine in (\ref{wtcc}), which  is maximized at $|r| =1$.
Here we have chosen the range of amplitudes $\bar A_\phi \leq 0.1$
which is the range where, in principle, the linear approximation is
arguable.  Notice that for expanding the sine function in
Eq.~(\ref{wtcc}), $2 \pi /\Phi_0 \delta \widetilde \phi \ll 1 $ which
implies $ \bar A_\phi \ll 0.1$.
To better quantify such a number, let us consider the case of a cavity
build up with JJs as mirrors.  Considering the quantization of the
photons inside the cavity $\phi = \sqrt { \hbar Z} (a^\dagger + a)$
yields that $2 n +1 = \bar A_\phi^2 \Phi_0^2/\hbar Z$.  This reasoning
implies that at amplitudes  $  \bar A_\phi  =0.1$ the number of
photons is around 4 (see also \cite{Peropadre2012}).
When considering the mirror formed by N junctions and recalling
Sect. \ref{sec:tunable} the jump at each junction is spread among the
junctions:
$\delta \phi_n = \delta \phi /N$ [Cf. Eq. (\ref{cNjj})].  Consequently the
number of photons scale at the mentioned amplitude $  \bar A_\phi
=0.1$  as $4 N^2$.

\section{Conclusions and outlook}

In summary, we have discussed the scattering characteristics of
Josephson junctions when they are embedded in superconducting
transmission lines.
The JJs are resonant scatterers such that, whenever the incident photon
matches the plasma frequency the junction behaves as a perfect
mirror.  Away from the resonance the junction is transparent.  
In this work we also included dissipative effects, which play a
role near resonance by degrading the perfect reflection.  
Importantly enough the broadband, resonance frequency and even the
dissipation can be tuned.  This is the great advantage.

The frequency dependence of the scattering can be used as a building
block for metamaterials, tailoring the photon propagation as discussed
previously \cite{Zueco2012}.
 In this
work, however, we have used this resonant character to propose the building
up arrays of coupled quantum cavities~\cite{Nori2010}. 
Importantly enough the coupling may be tuned in situ via external
fields.
We show that it is possible to use coupled-cavity array as a basis for
the study of polariton physics, for instance, bridging the gap between
the study of JJ quantum metamaterials and the quantum simulation of
many-body physics.

Finally we discussed the nonlinear corrections to the scattering.
While the calculations are expected to be valid in the few photon
limit we argue that the appearance of those nonlinear corrections can
also be controlled by the number of junctions forming the mirrors.
This can be used both for minimizing the nonlinear corrections or to
favor them for achieving nonlinear cavity-cavity coupling
\cite{Peropadre2012}.  The latter is important for the study of phases
in Bose-Hubbard like models \cite{Jin2013}.

This work was supported by Spanish government projects FIS2009-10061,
and FIS2011-25167 cofinanced by FEDER funds. We thanks Arag\'on
government support to group FENOL, CAM research consortium QUITEMAD
and PROMISCE European project.


\end{document}